\documentclass[a4paper]{article}

\usepackage{cite}
\usepackage{todonotes}
\usepackage{subcaption}
\usepackage{comment}
\usepackage{flushend}
\usepackage{dblfloatfix}
\usepackage{multirow}
\usepackage{threeparttable}

\usepackage{INTERSPEECH2022}

\title{Computing with Hypervectors for Efficient Speaker Identification}
\name{Ping-Chen Huang$^1$, Denis Kleyko$^{1,2}$, Jan M. Rabaey$^3$,  Bruno A. Olshausen$^1$, Pentti Kanerva$^1$}
\address{
  $^1$Redwood Center of Theoretical Neuroscience, University of California, Berkeley, USA\\
  $^2$Intelligent Systems Lab, Research Institutes of Sweden, Sweden\\  
  $^3$Berkeley Wireless Research Center, University of California, Berkeley, USA}

\email{\{pingchen.huang,denkle,jan\_rabaey,baolshausen,pkanerva\}@berkeley.edu}

\begin{document}

\maketitle
\begin{abstract}

We introduce a method to identify speakers by computing with
high-dimensional random vectors.  Its strengths are simplicity and
speed.  With only 1.02k active parameters and a 128-minute pass through
the training data we achieve Top-1 and Top-5 scores of 31\% and 52\%
on the VoxCeleb1 dataset of 1,251 speakers.  This is in contrast to
CNN models requiring several million parameters and orders of magnitude
higher computational complexity for only a 2$\times$ gain in
discriminative power as measured in mutual information.  An additional
92 seconds of training with Generalized Learning Vector Quantization (GLVQ)
raises the scores to 48\% and 67\%. A trained classifier classifies 1
second of speech in 5.7 ms. All processing was done on standard
CPU-based machines.

\end{abstract}
\vspace{.4em}

\section{Introduction}

With the emergence of Internet-of-Things devices, speaker recognition
at the edge is desirable as it can enable smart environments,
cyber-physical security, and robotic control, etc.  However, speaker
recognition is now done mostly in the cloud due to the constrained
resources and battery capacity of small devices that are unable to
run complex models.  Adapting to new speakers calls for
lightweight and efficient algorithms suitable for {\it on-device} and
{\it online} learning.

Developments in deep neural networks have led to end-to-end speaker recognition systems that achieve high accuracy on noisy and uncontrolled
speech data~\cite{nagrani2017voxceleb},~\cite{chung2018voxceleb2},~\cite{kim2021adaptive}.
Although neural networks have the ability to deal with noisy
real-world data, they are expensive to train due to iterative
back-propagation using gradient descent.  Moreover, the whole network
may need to be re-trained when adding new speakers.  Non-neural-network
approaches include the traditional Gaussian Mixture
Model--Support Vector Machine~\cite{campbell2006support}, and the more
recent Joint Factor Analysis~\cite{kenny2005joint} or the i-vector
approaches~\cite{dehak2009support}.  However, training these models usually requires running the
Expectation Maximization (EM) algorithm iteratively, which can also be
computationally intensive for large datasets.

We propose a speaker-recognition approach based on computing with
high-dimensional (HD) vectors, also called ``Hyperdimensional''~\cite{kanerva2009hyperdimensional}.
By mapping data to nearly
orthogonal vectors in a high-dimensional space and computing with
simple yet powerful operations on vectors, {\em the intrinsic structure of
the data can be revealed in a manner that is effective for classification}~\cite{wong2018negative,OsipovHyperSeed2021}.  HD computing has
provided an efficient way to analyze various types of
data~\cite{kleyko2021survey} and to achieve fast, online and
incremental learning in dealing with text~\cite{joshi2016language, AlonsoHyperEmbed2020},
multi-modal
bio-signals~\cite{rahimi2018efficient,moin2021wearable,zhou2021memory,ge2021seizure}, classifying spoken letters~\cite{imani2017voicehd}, and others~\cite{KleykoSurveyVSA2021Part2}.
This work extends the application of HD computing to speaker recognition.


We start by describing the idea of computing with hypervectors, and
the operations that are used to encode speech (Section~\ref{sec: overview}).  The proposed speech
encoder aims to capture the pronunciation variations between speakers
into a speaker profile hypervector (Section~\ref{sec:encoder}). The profile is computed in three
steps.  First, the formants in a time slice are encoded into a
hypervector, to capture the variation of the signal over frequencies.
Then hypervectors for consecutive time slices are encoded into an
$N$-gram hypervector, to capture the variation of the signal over
time. Finally the $N$-gram hypervectors of a speech sample are added
together, to form a profile hypervector that summarizes the course of the power spectrum over time.  Ways to improve this basic algorithm using GLVQ are also explored.

A metric is introduced to evaluate the efficiency of speaker identification systems in terms of the training energy per 1-bit of information gain.  This work has achieved a highly competitive energy efficiency due to its small number of active parameters during operation and one-shot learning.


\section{Computing with Hypervectors}
\label{sec: overview}

HD computing originates from Holographic Reduced
Representations in the early 1990s~\cite{plate1994distributed,
  plate2003HRRbook}. For speech processing, it provides a formulaic way to encode the frequency and temporal structure of a spectrogram into a fixed-dimensional vector.
  
``Hypervectors" refer to high-dimensional ($D >$
1,000) {\it seed vectors} and to vectors made from them with
three operations: {\it addition} ($+$), {\it multiplication} ($*$), and {\it
  permutation} ($\rho$). The seed vectors are chosen at random to
represent basic entities---they are like atoms from which 
everything else is built. Here 
they represent differences of spectral power in adjacent frequency
bins of a time slice. 
We use random bipolar vectors (of $+1$s and $-1$s) as seeds.
Addition and multiplication happen
coordinate-wise, and permutations reorder (shuffle) hypervector
coordinates. The {\it similarity} of vectors is measured with the
cosine, which equals 0 when the vectors are orthogonal---in a
high-dimensional space nearly all pairs of vectors are approximately
orthogonal.  Computing with vectors is like traditional
computing with numbers, except that addition and multiplication now
operate on vectors, and no arithmetic operation corresponds to the
permutation of coordinates.
  
Unlike most machine-learning methods that require iterative training,
HD computing offers one-shot and online learning.  Learning happens in
a single pass over samples from known speakers.  The pass produces a $D$-dimensional profile hypervector---a class prototype---for each speaker.
Profiles for test samples are made with the same algorithm and are
identified with the most similar speaker profiles. Therefore, the
model need not be retrained when speakers are added.

\section{Encoding Speech}
\label{sec:encoder}

Learning the statistics of a signal over time is particularly natural and efficient with hypervectors.  The proposed encoder aims to capture the unique pronounciation variation between speakers,  similar to the phone $N$-gram-based modeling~\cite{kohler2001phonetic}.  In this approach,  a profile hypervector is designed to learn the unique course of the formants over time for each speaker.  

\begin{figure}[t]
  \centering
  \includegraphics[width=1\linewidth]{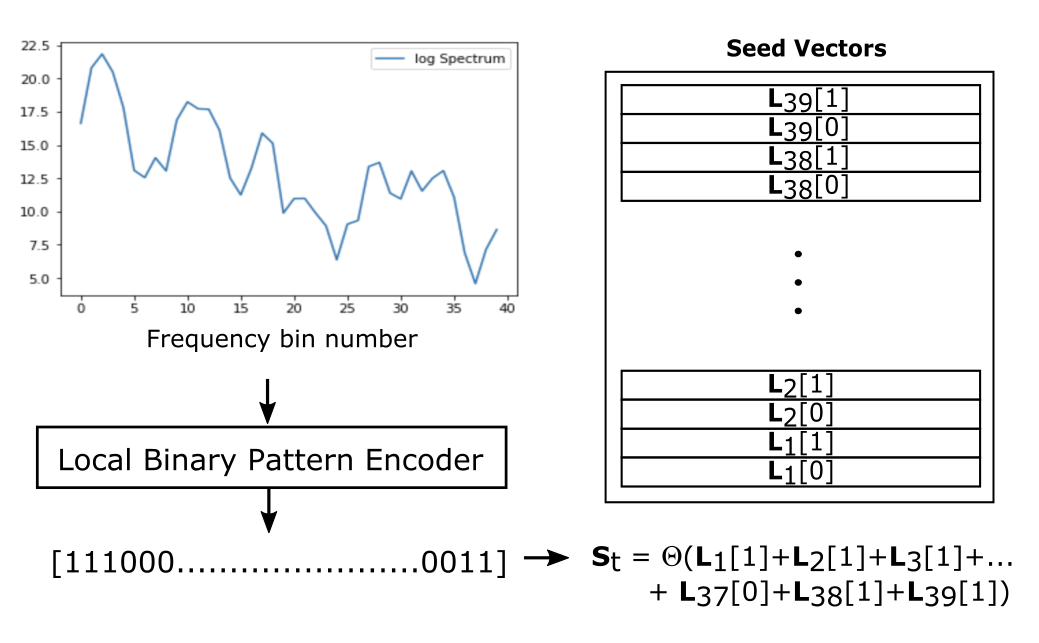}
  \caption{Encoding a spectrum at time $t$ into a hypervector $\mathbf{S}_t$. The spectrum  represents 5 ms of speech sampled at 16 kHz.}
  \label{fig:encoder}
\end{figure}

In a spectrogram,  typically up to 4 formants stand out at any moment
of time. Taking a spectrum a slice at a time,  for example the upper left plot in Figure~\ref{fig:encoder},  a formant can be identified by rising power to the left of it and falling to the right.  Therefore,  formants can be located by comparing the power in adjacent bins.  A simple local binary pattern (LBP) encoding~\cite{burrello2019hyperdimensional} is used here to encode the locations of the formants in a time slice.  From the first bin to the second bin,  the LBP encoder looks at whether the power increases or remains the same, or decreases, and reports 1 or 0.  In the VoxCeleb dataset, audio is sampled at 16 kHz, and so the first 40 bins of a power spectrum computed over a 5-ms window cover frequencies from 0 to 8,000 Hz.  The power in 40 bins gives rise to 39 differences between neighboring bins, resulting in a 39-bit LBP.

The hypervector $\mathbf{S}_t$ for the spectrum at time $t$ summarizes the output of the LBP encoder.  It is made from bipolar seed
vectors that represent the 0s and the 1s of the LBP. There are a total
of 78 seed vectors to choose from, corresponding to power going up or
down at each bin. For example,  $\mathbf{L}_i[1]$ or $\mathbf{L}_i[0]$ represents the power in the ($i+1$)-th bin greater or less than that in the $i$-th bin.  Then according to the output of the LBP encoder,  39 seed vectors are selected and added together.  Finally,  the resulting sum vector is transformed to a bipolar vector of 1s and $-1$s by thresholding it at zero:
\begin{equation}
\mathbf{S}_t = \theta\left(\sum_{i=0}^{39} \mathbf{L}_i[1;0]\right)
  \label{eq1}
\end{equation}
where
\begin{equation}
\theta (x) =
    \begin{cases}
      1 & \text{if $x>0$ }\\
      -1 & \text{if $x<0$ }\\
    \end{cases}       
  \label{eq2}
\end{equation}
is applied coordinate-wise.
Only looking at whether the power is increasing or decreasing has the advantage of not being affected by the $1/f$ power characteristics and the loudness of speech. 
It is important to note that this encoding yields similar hypervectors for similarly located formants.

Hypervectors $\mathbf{S}_t$ for individual spectra are combined into spectra over time by encoding them in $N$-grams and adding into a profile hypervector. Empirically, we found that trigrams and tetragrams work the best.  Sampling of spectrum slices at 20-ms intervals means that these $N$-grams represent 60--80 ms of speech, or approximately the length of a phoneme.  

A trigram vector is made by permuting the vector for the first spectrum twice, permuting the vector for the second spectrum once, taking the third as is,  and multiplying the three vectors coordinate-wise,  i.e.:
\begin{equation}
\text{trigram}[t]=\rho^2 \mathbf{S}_{t-2}*\rho \mathbf{S}_{t-1}*\mathbf{S}_t
  \label{eq3}
\end{equation}
where $\rho$ denotes permuting the vector once,  and $\rho^2$ denotes permuting the vector twice.  Permutation was implemented by random shuffling the indices of coordinates of the vector.  

Finally, the profile hypervector is formed by summing the trigram vectors over time within an utterance,  or across multiple utterances. 
In VoxCeleb dataset,  utterances have been collected for each speaker in different contexts with different recording quality and background noise. 
As it will become clear in Section~\ref{sec:results},  two kinds of profile hypervectors were generated for each speaker: (1) for each context/video/subfolder and (2) one across all the speaker's contexts.  For speaker $s$,  his/her context profile hypervector $\mathbf{V}_{s,c}$ for context $c$ is the sum of all trigrams from all utterances in that context,  i.e. $\mathbf{V}_{s,c}=\sum_t\text{trigram}_{s,c}[t]$,  and the final profile hypervector for speaker $s$ is $\mathbf{V}_s=\sum_{c}\mathbf{V}_{s,c}$
%
%

\begin{figure}
     \centering
     \begin{subfigure}[b]{0.23\textwidth}
         \centering
         \includegraphics[width=\textwidth]{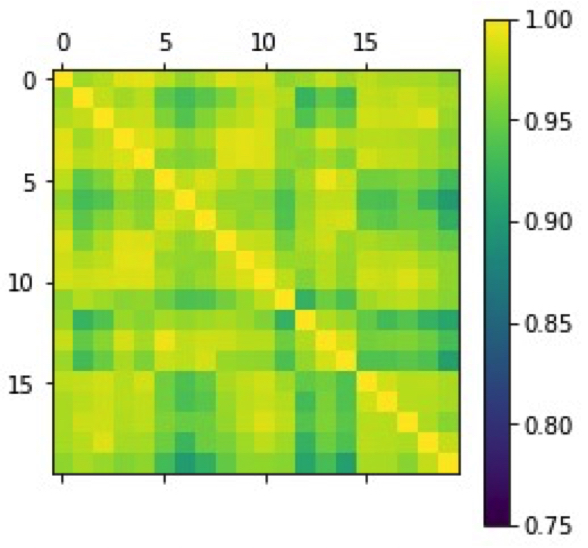}
         \caption{Without weighting}
         \label{fig:3a}
     \end{subfigure}
     \hfill
     \begin{subfigure}[b]{0.23\textwidth}
         \centering
         \includegraphics[width=\textwidth]{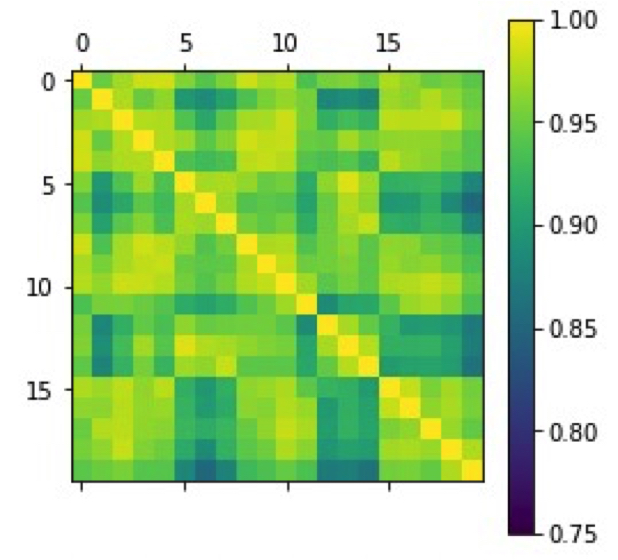}
         \caption{With weighting}
         \label{fig:3b}
     \end{subfigure}

        \caption{Correlation matrices of the first 20 speakers' profile hypervectors.}
        \label{fig:correlation}
\end{figure}

\section{Experiments}
\label{sec: exp}
\subsection{Experimental Setup}

We used the VoxCeleb dataset~\cite{nagrani2017voxceleb} to develop and
test the algorithm.  It consists of speech from 1,251 speakers
collected from YouTube videos. Each speaker has a folder, divided into
a number of subfolders that contain the audio files.  The subfolders
come from different videos and are referred to as different
``contexts.''  The speech files in the subfolders are called
``utterances.''  Following the same procedure as
in~\cite{nagrani2017voxceleb}, we reserved one subfolder/context for testing that had
the fewest utterances but at least five. Each speaker is trained with 904 seconds of speech and tested with 64 seconds, on the average.
Top-1, Top-5, and Top-10 accuracies were calculated.

\subsection{Input Features}
Spectrograms are computed using a 5-ms Hann window and a step size of 20 ms. A short window---5 ms vs, the commonly used 25 ms---simplifies the LBP encoding of the formants by smoothing over the  multiples of the fundamental frequency and its overtones.  Spectrum slices are sampled at 20-ms intervals such that a trigram or a tetragram represents 60--80 ms of speech,  approximately the length of a phoneme.  

\subsection{Encoder Design}
\label{sec:design}
Before training and testing on the entire dataset,  a few parameters of the encoder need to be determined,  such as the dimension of the hypervectors,  which $N$-gram to use,  and the number of frequency bins to encode.  The dimension of the hypervectors $D$ was chosen to be 1,024.  To determine which $N$-gram to use,  uni-gram to penta-gram were used in the encoder to train and test on 40 speakers' data,  which is sufficient to indicate the results for the entire dataset. 
The results suggested that tetragrams and trigrams are comparable and perform better than the others.  Therefore,  trigrams were chosen for the encoder.  Similarly,  different number of frequency bins (26, 32, and 40) were encoded on 40 speakers' data,  and results suggested that encoding the full number of bins (40) performs best.  
For the proposed approach, the training and testing on 40 speakers' data take roughly 5 minutes on an Apple M1 processor,  so one can quickly test design parameters.

\subsubsection{Weighting}
Voice activity detection is usually necessary so that desirable features are extracted only from speech segments. The algorithm so far treated a period of silence the same as speech. To counter the lack of information in silence or a weak signal, we use the total energy in the spectrum to weight its hypervector before including it in the $N$-gram.  The total energy of a spectrum slice at time $t$ is $E_t=\sum_f |x_f|^2$.  Thus the weighted trigram at time $t$ is
\begin{equation}
(\rho^2 \mathbf{S}_{t-2}*\rho \mathbf{S}_{t-1}*\mathbf{S}_t)\cdot(E_{t-2}E_{t-1}E_t)^\alpha
  \label{eq4}
\end{equation}
where the exponent $\alpha$ was determined empirically as described in Section~\ref{sec:design}.  We found that weighting with the 0.3  power of energy works well with trigrams.  Figures~\ref{fig:3a} and~\ref{fig:3b} show the correlation matrices of the first 20 speakers' profile hypervectors before and after applying the weights.  It can be seen that the correlations drops after applying the weighting.  
\subsubsection{Normalized Weights}
\begin{figure}[t]
  \centering
  \includegraphics[width=0.9\linewidth]{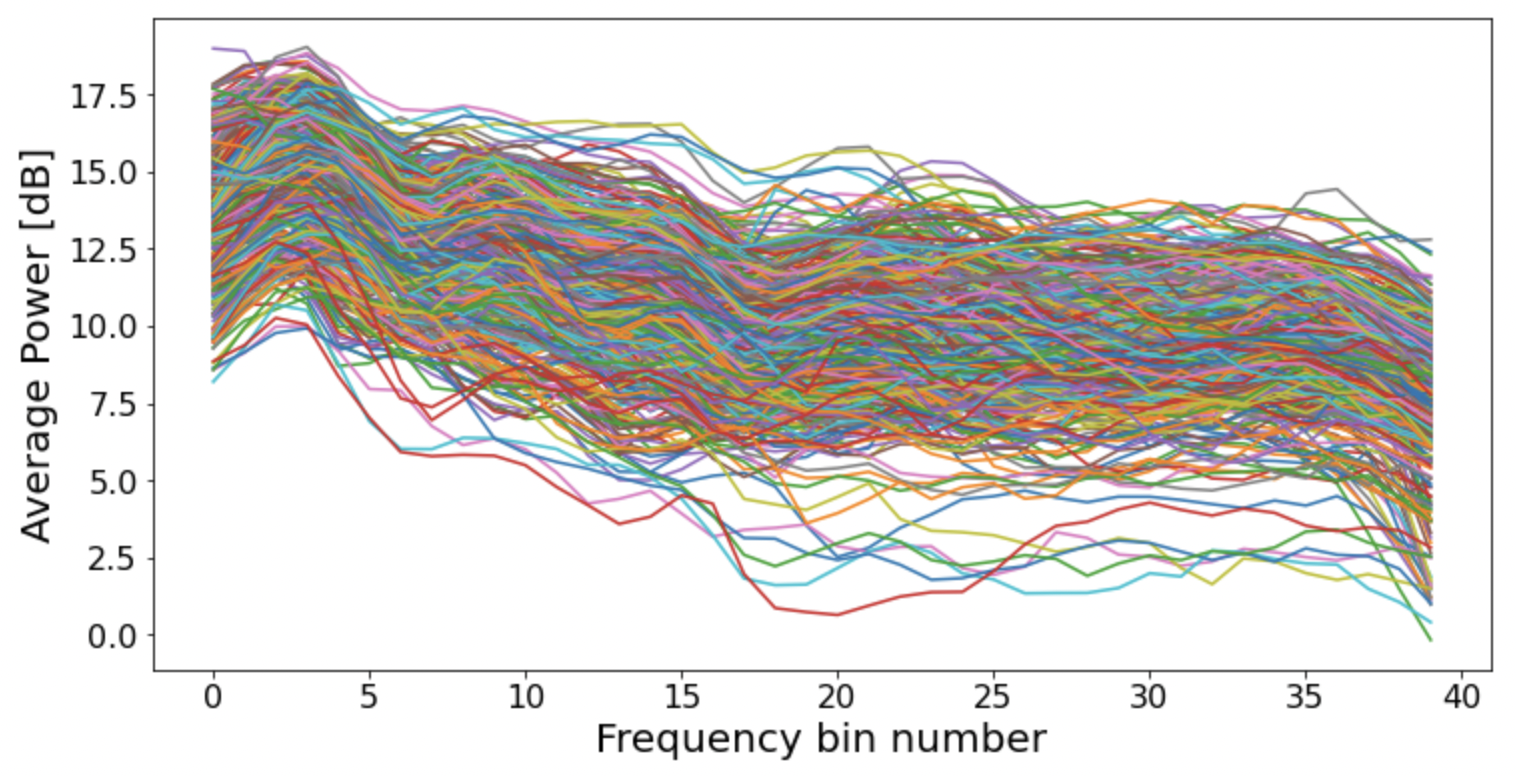}
  \caption{Average power per utterance for a speaker across multiple contexts from the VoxCeleb dataset. Different colored lines denote different utterances.}
  \label{fig:pave}
\end{figure}

Performance may be further improved by normalizing the weights to discount large variations in the power of speech segments across different contexts (due to lack of control over recording conditions in the VoxCeleb dataset).  As shown in Figure~\ref{fig:pave},  the average power per frequency bin over an utterance for the same speaker can vary over 15 dB.  To avoid favoring contexts with louder voices or even noise,  the weight for the hypervector of each spectrum slice is scaled by the ratio between the desired maximum bin power and a speaker's maximum bin power averaged over the utterance where the hypervector at time $t$ is computed from:
\begin{equation}
c_t=\frac{P_{\text{target}}}{P_{\text{max, utterance}}}
  \label{eq5}
\end{equation}
where $P_{\text{target}}$ is the targeted max bin power set for all utterances to be adjusted to and can be an arbitrary constant.  We set it to the average of the largest bin power over the first 40 speakers.  $P_{\text{max, utterance}}$ is the maximum bin power of the utterance being considered and is computed during training.  This essentially makes all utterances' maximum bin power equal,  so the weights no longer favor contexts with higher spectral power.  Applying this normalizing ratio leads to an increase of 4.3\% and 7.3\% in the Top-1 and Top-5 accuracies from the weighted case.

\begin{table*}[th]
  \caption{Identification accuracy on VoxCelb1 dataset using the proposed methods}
  \label{tab:summary}
  \centering
  \begin{tabular}{ c c c c c c}
    \toprule
    \multicolumn{1}{c}{\textbf{Method}} &
     \multicolumn{1}{c}{\textbf{Weighting applied to $\mathbf{S}_t$}} &       \multicolumn{1}{c}{\textbf{Further processing}} &
\multicolumn{1}{c}{\textbf{Top-1}} &       \multicolumn{1}{c}{\textbf{Top-5}} &   \multicolumn{1}{c}{\textbf{Top-10}} \\

    \midrule
     Baseline                       &none & none & $0.249$ & $0.414$  & $0.500$~~~           \\
     Weighting                   &$E_t^{\alpha}$  & none   & $0.270$ & $0.447$  & $0.557$~~~           \\
    Normalized Weighting & $(c_tE_t)^{\alpha}$ & none  & $0.313$ & $0.520$  & $0.613$~~~           \\
Refinement of profile hypervectors        & $(c_tE_t)^{\alpha}$ & GLVQ  & $0.479$ &$0.674$  &$0.755$~~~           \\
    \bottomrule
  \end{tabular}
  \end{table*}

\begin{table*}[th]\footnotesize
  \caption{Comparison with prior works on VoxCeleb1 dataset for speaker identification.}
  \label{tab:comparison}
  \centering
\begin{threeparttable}
\begin{tabular}{ c c c c c c c c }
    \toprule
    \textbf{Method} &  \textbf{Top-1}  & \textbf{Top-5} &
    \textbf{Active Parameters\tnote{1}} & \textbf{Stored} &
    \textbf{Mutual} & \textbf{Training} &
    \textbf{Classification} \\
    & & & \textbf{during training}& \textbf{Parameters} & \textbf{Info.} &\textbf{Method (Time)} &\textbf{Speed} \\
    \midrule
     i-vectors/SVM~\cite{nagrani2017voxceleb} & $0.490$ & $0.566$ &  not reported & 1M\tnote{2}  &  4.04 bits& iterative EM+SGD\tnote{3} & not reported\\
         \midrule
i-vectors/PLDA/SVM~\cite{nagrani2017voxceleb} & $0.608$ & $0.756$ & not reported & 0.5M\tnote{4} &5.29 bits & iterative EM+SGD & not reported\\
    \midrule
CNN~\cite{nagrani2017voxceleb} & 0.805 & 0.921 & 67M & 67M & 7.57 bits & iterative SGD & not reported\\
    \midrule
ACNN~\cite{kim2021adaptive} & 0.855 & 0.953 & 4.69M & 4.69M & 8.20 bits & iterative SGD & not reported\\
    \midrule
\multirow{2}{*}{\textbf{This work}} & \multirow{2}{*}{$\mathbf{0.479}$} & \multirow{2}{*}{$\mathbf{0.674}$} & HD: 1.02k & HD: 1.28M & 
\multirow{2}{*}{3.93 bits} & HD: one-shot (128 min) & 5.7 ms per\\
& & & GLVQ: 2.05k & GLVQ: 1.28M & & GLVQ: SGD (92 sec) & 1-sec test sample\\
\bottomrule
  \end{tabular}
  \begin{tablenotes}\footnotesize
\item[1] Number of parameters updated for a single pass of one data sample.
\item[2] Assuming two 400-dimensional vectors~\cite{nagrani2017voxceleb} were stored for each speaker's SVM.
\item[3] Generally i-vector extractor is trained with the EM algorithm,  and SVM is trained with the SGD algorithm.
\item[4] Assuming two 200-dimensional vectors~\cite{nagrani2017voxceleb} were stored for each speaker's SVM.
\item[5] SVM: Support Vector Machine; PLDA: Probabilistic Linear Discriminant Analysis; CNN: Convolutional Neural Network; ACNN: Adaptive CNN; EM: Expectation Maximization; SGD: Stochastic Gradient Descent.
\end{tablenotes}
\end{threeparttable}
  \end{table*}

\subsection{Refinement of profile hypervectors with \\ 
Learning Vector Quantization}
\label{sec:lvq}

The profile hypervector $\mathbf{V}_s$ that sums a speaker's context hypervectors corresponds to centroid-based classification, which is commonly used in speech and signal processing (e.g.,~\cite{RasanenMultivariate2015,KleykoTradeoffs2018, GeClassificationReview2020}) due to its simplicity, although it does not guarantee the most accurate classification
~\cite{RosatoHDDistributed2021,KarlgrenSemantics2021}.
Of the various classifiers that can take hypervectors as input (e.g.,~\cite{RachkovskijClassifiers2007,RahimiBiosignal2016,imani2017voicehd,KleykoDensityEncoding2020,DiaoGLVQHD2021}), we used the Generalized Learning Vector Quantization (GLVQ)~\cite{SatoGLVQ1995},  since it is natural to use context hypervectors to initialize prototypes and then refine them  iteratively~\cite{DiaoGLVQHD2021}.
In each iteration,  the classifier uses one misclassified context hypervector $\mathbf{V}_{s,c}$ from one speaker to update the speaker's profile vector $\mathbf{V}_s$ (``prototype'') as well as the profile vector of the nearest (the most similar) speaker.
In this manner, the classification accuracy improves after each iteration.

\subsection{Results}
\label{sec:results}

Table~\ref{tab:summary} summarizes the identification accuracies on the entire dataset using the methods described above. The baseline is based on a very simple encoder, and any reasonable feature engineering keeps improving the score.  Most importantly,  training and testing on the entire dataset require only 1.02k active parameters and take only 135.6 minutes (training: 128 minutes; testing: 7.6 minutes) on a regular CPU-based Linux machine (Intel Xeon\textregistered~CPU @ 2.40GHz) with a maximum usage of 5 cores during the program.  The classification speed is 5.7 ms per 1-second of speech.

\begin{figure}
    \centering
    \includegraphics[width=0.9\columnwidth]{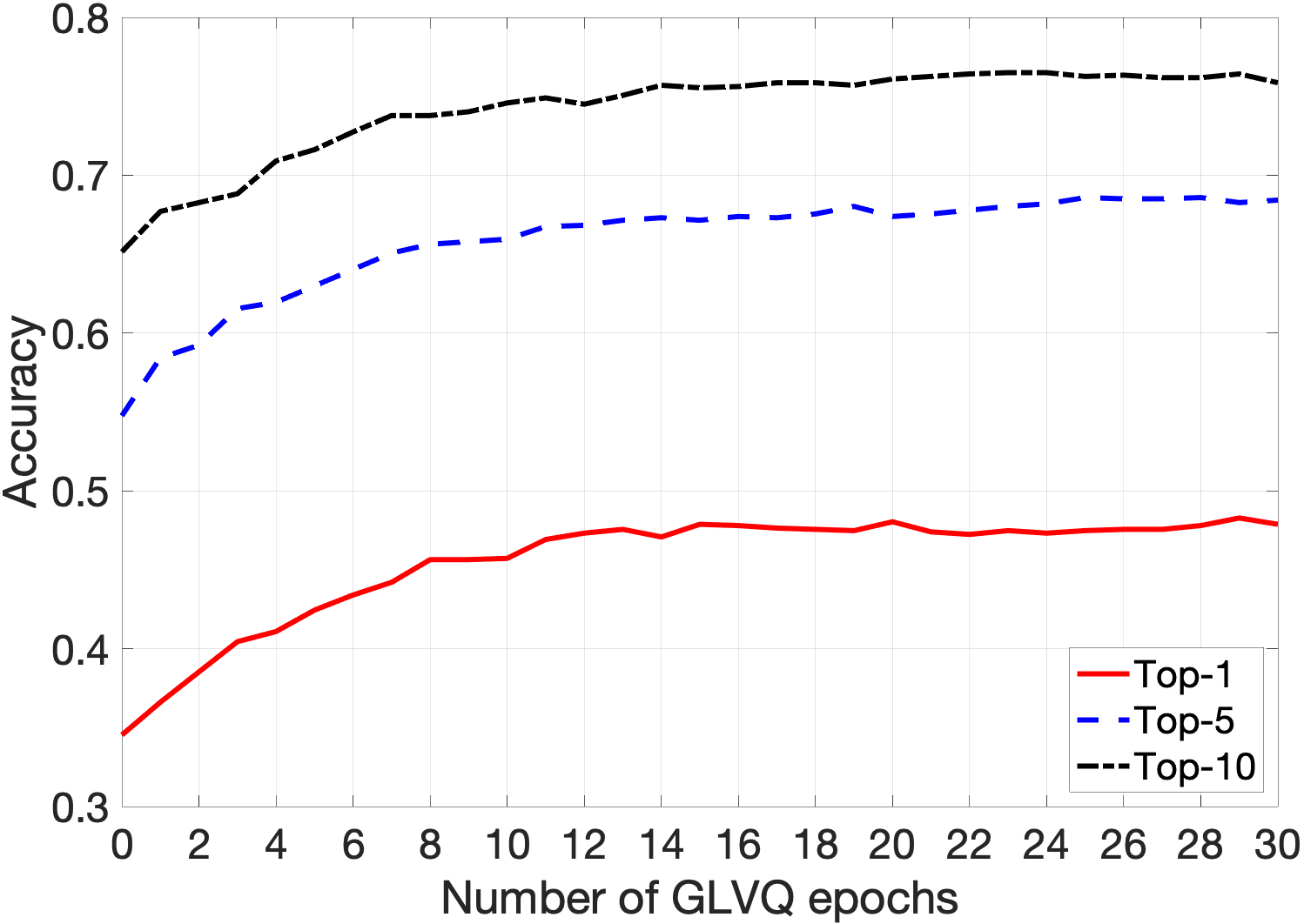}
    \caption{The test accuracy of the GLVQ classifier against the number of training epochs.
    }
    \label{fig:GLVQ:acc}
\end{figure}

The GLVQ classifier with a single prototype per speaker is used to obtain refined profile vectors for speakers from the context hypervectors.  After every epoch (i.e., a pass over all training samples) the classification accuracies were evaluated. Figure~\ref{fig:GLVQ:acc} shows the results for Top-1, Top-5, and Top-10 test accuracies.  Epoch $0$ corresponds to the accuracy of the centroid-based classification. 
For example, in just two epochs Top-1 accuracy increased from $0.313$ to $0.385$, 
and the accuracy started to saturate after approximately $15$ epochs reaching 
$0.479$, $0.674$, and $0.755$ as Top-1, Top-5, and Top-10 accuracies, respectively.  Running GLVQ for 30 epochs took only 92 seconds on a regular CPU-based laptop.

Table~\ref{tab:comparison} compares this work to the state-of-the-art non-neural network and neural network approaches.  Although identification accuracy has been the only metric reported for most methods,  the cost that comes with it is also an important factor that cannot be ignored.  Therefore,  we proposed a metric to evaluate the efficiency of a design: the energy to train a model per 1-bit of information gain,  i.e.:
\begin{equation}
\frac{E_{\text{train}}}{I} = \frac{\text{(\# active paramaters during training)}\cdot t_{\text{train}}}{I}
  \label{eq6}
\end{equation}
where $t_{\text{train}}$ is the training time,  and $I$ is the information gain (in bits) from the speaker identification system. \footnote{$I$ can be estimated from the Top-1 accuracy $p$ as $I=H(Y)-H(Y|X)=log_21251-\left( p\cdot log_2\frac{1}{p} + (1-p)\cdot log_2\frac{1250}{1-p} \right)$,  where $X$ and $Y$ are the input speaker id and output speaker id of the identification system, assuming every speaker is equally likely to appear at the input and has the same probability $p$ to get identified at the output. If not identified, a speaker gets misclassified to any one of the rest 1250 speakers equally likely.}
Although neural networks achieve 4 more bits of information gain,  the energy to train the network is much larger than 2 times,  as they take more than 4500$\times$ active parameters to train iteratively over an unspecified number of epochs.  For i-vector approaches,  they also require iterative training for i-vector extraction and the per-speaker SVM training.  Due to the limited information reported from other works,  we are not able to quantify their efficiency. Considering relatively few active parameters used during training and the one-shot learning algorithm,  we believe that our approach leads to highly energy-efficient systems for speech processing.

\section{Discussion}
In this work,  we have studied the application of a new computing paradigm to the encoding of speech.  With a simple encoding scheme and reasonable feature engineering,  it has achieved highly competitive efficiency for its information gain.  The results obtained so far are solely based on making use of one acoustic feature (formants) and their course over a short time.  There are many more acoustic features yet to be considered,  such as the pitch and cepstral coefficients.  HD computing is especially suited for encoding a combination of features and producing a fixed-dimensional representation for them~\cite{KarlgrenSemantics2021}.  Therefore,  its identification accuracy is expected to keep improving when combined with other acoustic
features, with a modest increase in computing time and memory use. This work can help to originate a simpler, more energy-efficient machine learning for speech processing.

\section{Acknowledgements}


PCH was supported by NSF ECCS-2147640.
PCH, DK, JMP, BAO, and PK were supported in part by the DARPA's AIE (HyDDENN Project).
DK, BAO, and PK were also supported in part by AFOSR FA9550-19-1-0241.
DK was supported by the EU's MSCA Fellowship (839179).

\newpage

\bibliographystyle{IEEEtran}

\bibliography{Bibliography_etal}

\end{document}